\documentclass[aps, a4paper,prl, twocolumn, amsmath]{revtex4}
\usepackage[T1]{fontenc}
\usepackage[utf8]{inputenc}
\usepackage[english]{babel}
\usepackage{amsmath}
\usepackage{amssymb}
\usepackage{graphicx}
\usepackage{caption}
\usepackage{lipsum}

\newcommand{\beq}{\begin{equation}}
\newcommand{\eeq}{\end{equation}}

\newcommand{\beqa}{\begin{eqnarray}}
\newcommand{\eeqa}{\end{eqnarray}}

\newcommand{\nn}{\nonumber}

\begin{document}
\title{\Large You are in a drawdown. When should you start worrying?}
\author{Adam Rej, Philip Seager \& Jean-Philippe Bouchaud}
\address{Capital Fund Management, 23 rue de l'Universit\'e, 75007 Paris}
\begin{abstract}
\it 
\center
{Trading strategies that were profitable in the past often degrade with time. Since unlucky streaks can also hit ``healthy'' strategies, how can one detect that something truly worrying is happening? 
It is intuitive that a drawdown that lasts too long or one that is too deep should lead to a downward revision of the assumed Sharpe ratio of the strategy. In this note, we give a quantitative answer to this 
question based on the exact probability distributions for the length and depth of the last drawdown for upward drifting Brownian motions. We also point out that both managers and investors tend to underestimate the length and depth of drawdowns consistent with the Sharpe ratio of the underlying strategy. 
}
\end{abstract}
\maketitle

\section{Introduction}
The search for automated trading strategies lies at the heart of quantitative trading. This is a complex process. First, caution should be exercised to avoid taking spurious statistical relationships for viable strategies (the so-called data snooping problem). Second, even if one believes the relationship to be valid, one has to determine the amount of ``in-sampleness'' or overfitting that has been employed in the process of looking for the strategy in question. Finally, a trading cost analysis usually leads to a slowed-down, lower-Sharpe version of the strategy. In practice, however, this is not the end of the story. The strategy that has made it into production should be monitored and its performance should be benchmarked against what one believes is the best estimate of the out-of-sample Sharpe ratio, with and without costs.  

Let us assume that a manager put into production a strategy for which she believes that the best-estimate before-cost Sharpe ratio is $\textrm{SR}^*$. After some time an unwelcome event sets in: the gross-of-cost PnL is exhibiting a significant drawdown. This may either be a drawdown that is particularly deep or one that lasts for long time (or both). The manager (or the investor!) starts asking herself whether she was right about her best-estimate Sharpe ratio. When should she start worrying? In what follows, we shall work out how the length and the depth of the last drawdown affect the best-estimate Sharpe ratio.

\section{The setup} \label{sec:setup}

Let us assume that the (log-)PnL of the strategy is well modelled by a drifted Brownian motion
\beq \label{process}
{\rm d} \textrm{PnL} = \mu\, {\rm d}t + \sigma\,{\rm d}W 
\eeq
over a finite time interval $(0,T)$. In what follows we will normalize the risk such that $\sigma=1$; the corresponding annualized Sharpe ratio is then $\textrm{SR} = \mu$. We measure time in years. It is a well-known property of the Brownian motion that, as long as the interval $(0,T)$ is finite, the maximum 
\beq
	{\textrm{PnL}}_{\text{max}} = \textrm{max}_{0 \leq s \leq T} \textrm{PnL}_s
\eeq
is well-defined and the last time the maximum occurred, $s_\textrm{max}$, is almost surely unique, and defines the ``length'' of the drawdown, $\ell := T - s_\textrm{max}$. 
The ``depth'' of the drawdown $d > 0$ is simply defined as the difference
\beq
	d := {\textrm{PnL}}_{\text{max}} - \textrm{PnL}_T\,.
\eeq
We illustrate these quantities in Figure \ref{fig:scheme}.
In the appendix we compute the unconditional probability distribution of the length of the last drawdown $\rho(\ell|{\textrm{SR}})$ and the probability distribution of the depth of the last drawdown $\psi(d|{\textrm{SR}})$. These densities depend on the Sharpe ratio of the process \eqref{process}, as well as the total time $T$.  

\begin{figure}
	\begin{center}
	\includegraphics[scale=0.25]{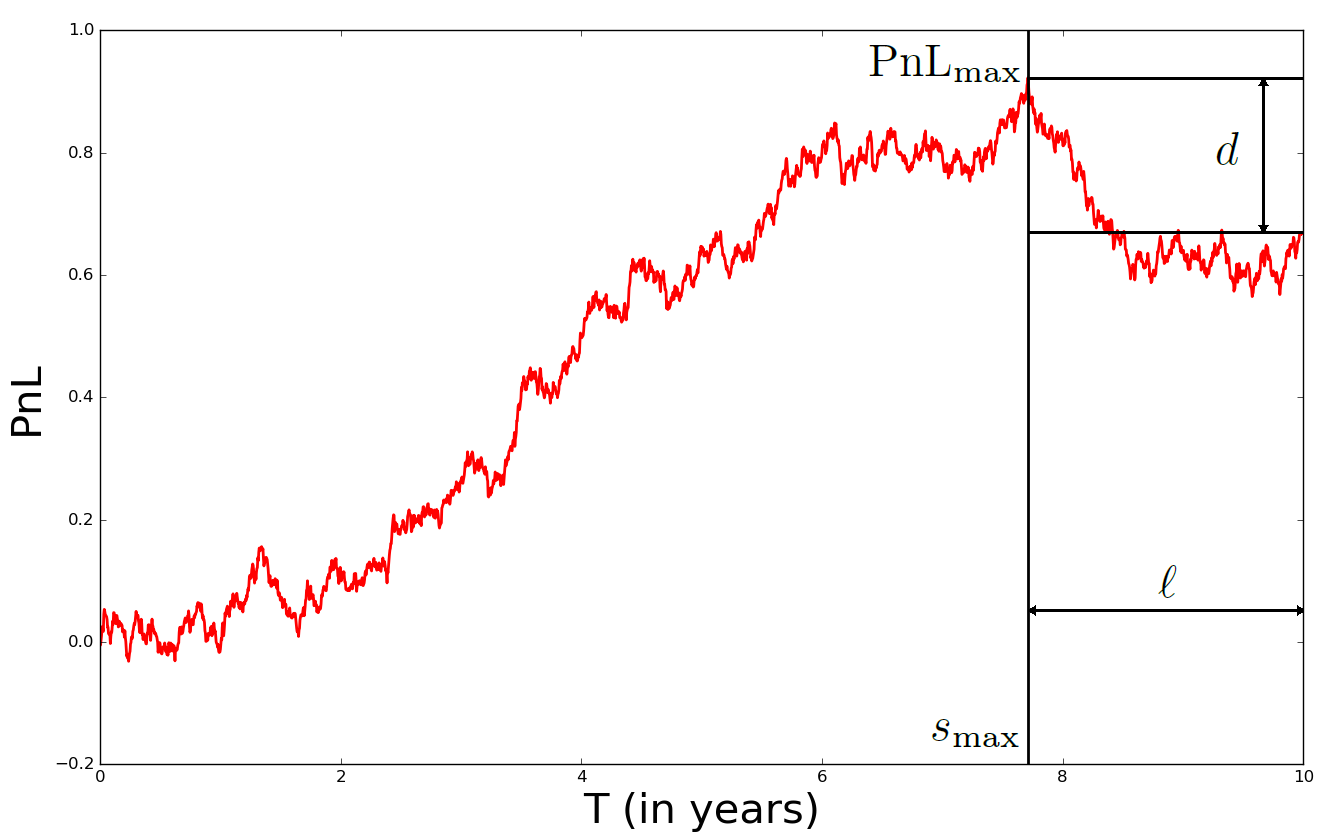}
	\caption{Schematic representation of a depth and a length of a drawdown for a 10y process assuming 257 trading days per year.} \label{fig:scheme}
	\end{center}
	
\end{figure}
\section{The test} \label{sec:test}
\begin{figure}
	\begin{center}
	\includegraphics[scale=0.24]{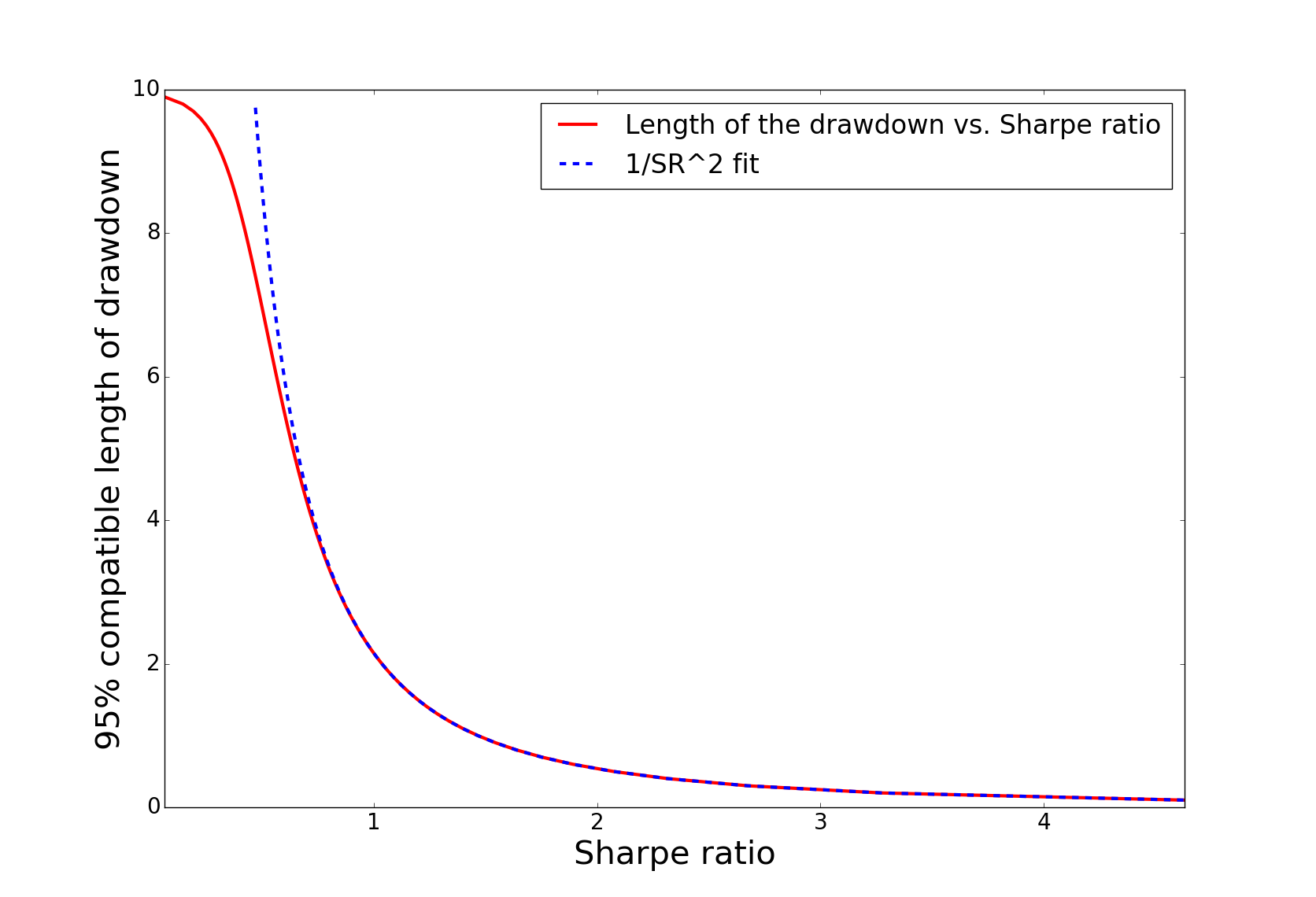}	
	\caption{The $5\%$ length (in years) of a drawdown as a function of the Sharpe ratio for a ten-year process, and a fit as $\ell_{5\%} = 2.14 \times \textrm{SR}^{-2}$.} \label{fig:conf}
	\end{center}
	\begin{center}
	\includegraphics[scale=0.23]{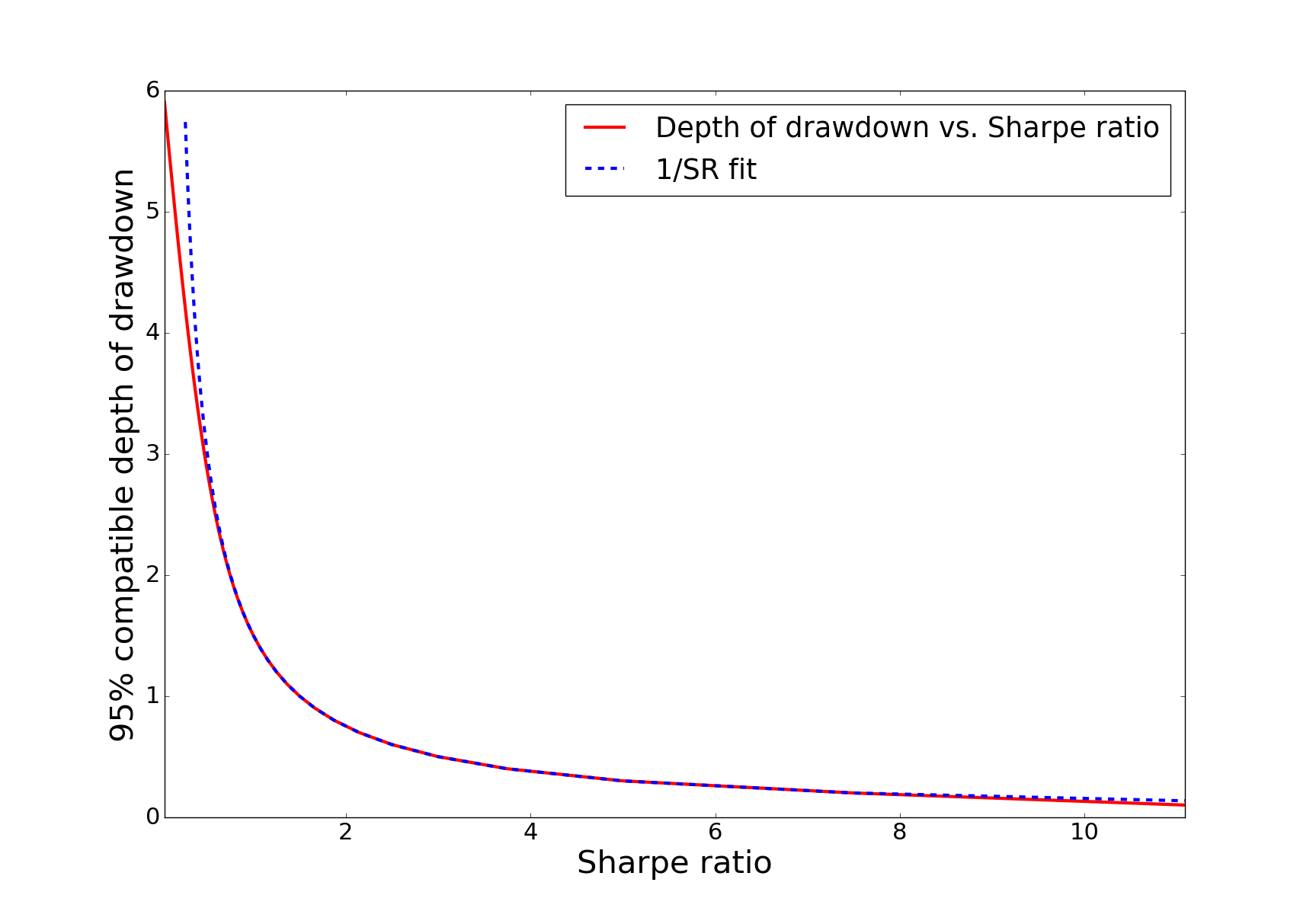}	
	\caption{The normalized $5\%$ depth of the drawdown interval as a function of the Sharpe ratio for a ten-year process, and a fit as $d_{5\%} = 1.50 \times \textrm{SR}^{-1}$.} \label{fig:draw}
	\end{center}
\end{figure}

Having these two densities at her disposal, can the asset manager test her Sharpe ratio hypothesis? 
Let us assume that she plots her PnL and determines that her strategy is in a drawdown of depth $d$ and length $\ell$ \cite{sigma}. We define $5 \%$ extreme drawdown length, conditional on the assumed Sharpe ratio $\textrm{SR}^*$, as
\beq \label{t95}
\int^{T}_{\ell_{5\%}} {\rm d}u\, \rho(u |\textrm{SR}^*) = 0.05\,,
\eeq
and the $d_{5\%}$
\beq \label{d95}
\int_{d_{5\%}}^{\infty} {\rm d}u\, \psi(u | \textrm{SR}^*) = 0.05\,.
\eeq
Both quantities have very intuitive interpretation. For a strategy with a Sharpe ratio $\textrm{SR}^*$ there is only 5\% chance that the the drawdown will last longer than $\ell_{5\%}$. In the same vein, there is 5\% chance that the depth of the drawdown will exceed $d_{5\%}$. We show the numerical dependence of $\ell_{5\%}$ and $d_{5\%}$ on the Sharpe ratio in Figures \ref{fig:conf} and \ref{fig:draw}. These results are consistent with the well-known fact (see e.g. \cite{book}) that for a Brownian motion the depth of a typical drawdown is inversely related to the Sharpe ratio, while the length of drawdown is inversely related to the {\it square} of the Sharpe ratio. We fitted and overlaid this heuristic dependence upon the 5\% curves. We observe almost perfect match for a wide range of values of Sharpe ratios. The mismatch for small Sharpe ratios is due to the finite time interval effects.

\begin{figure}
	\begin{center}
	\includegraphics[scale=0.12]{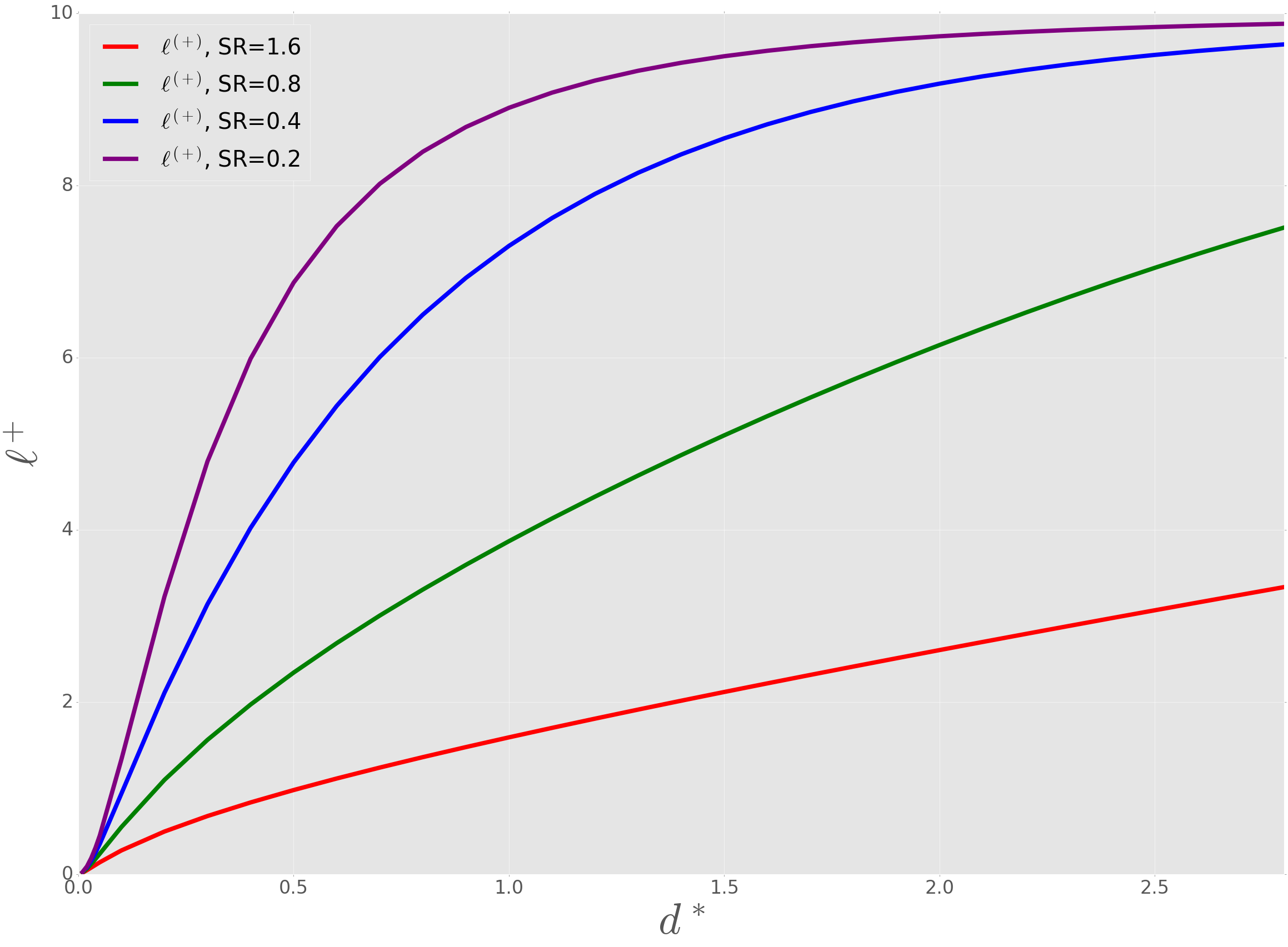}	
	\caption{The upper boundary of the conditional drawdown corridor, defined as $\mathbb{P}(\ell \geq \ell^{+} \,|\, d^*) = 0.05$, for different values of the Sharpe ratio of a 10y process. $\ell^{+}$ is expressed in years.} \label{fig:LD1}
	\end{center}
	\begin{center}
	\includegraphics[scale=0.12]{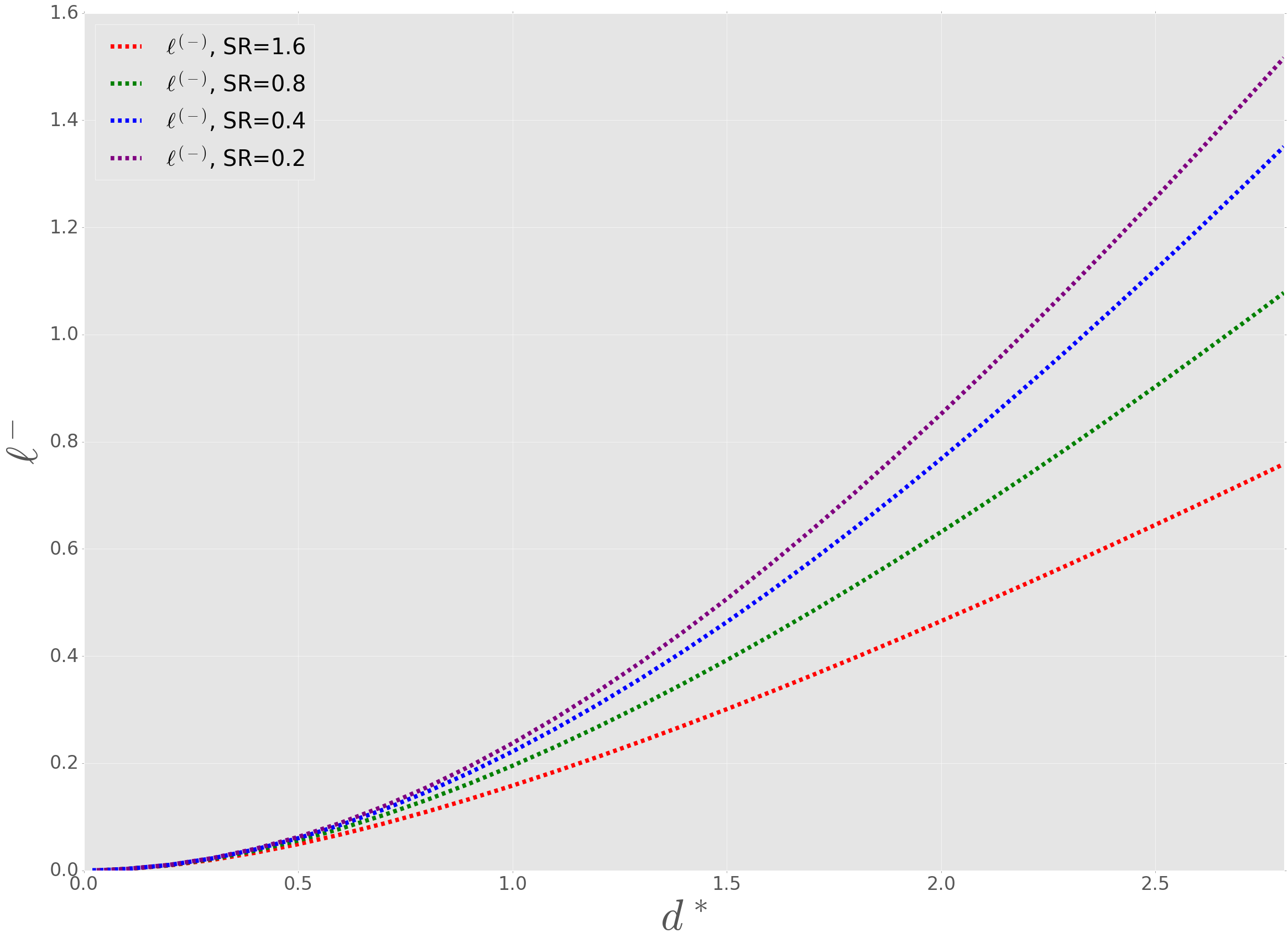}	
	\caption{The lower boundary of the conditional drawdown corridor, defined as $\mathbb{P}(\ell \leq \ell^{-} \,|\, d^*) = 0.05$, for different values of the Sharpe ratio of a 10y process. $\ell^{-}$ is expressed in years.} \label{fig:LD2}
	\end{center}
\end{figure}
Now the asset manager may compare the values of her current drawdown length $\ell$ and depth $d$ with the above 5\% values. If either of them lies outside its 5\% interval, what should the manager conclude? 
There are essentially three possible scenarios. 
\begin{itemize}
\item Scenario 1: Both the Brownian model and the assumption made about the Sharpe ratio $\textrm{SR}^*$ are acceptable. The deep or long drawdown observed is just bad luck or, put differently, a 5\% event. 
\item Scenario 2: The Brownian model is acceptable but the Sharpe ratio $\textrm{SR}^*$ has been overestimated.
\item Scenario 3: The Brownian model is inadequate and underestimates the probability of drawdowns. This can happen for various reasons like a non-Gaussian distribution of real-life returns, non-stationarity of the volatility, or weak positive autocorrelation of the returns of the strategy, etc. \cite{fracBrown} All these effects generate longer and/or deeper drawdowns.
\end{itemize}

The last scenario is of course important to keep in mind, but we argue that from a practitioner's point of view, it is always better to err on the side of caution, so that the Brownian model sets a very useful benchmark. The test we propose clearly does not allow to distinguish between the first two scenarios. It is thus possible that this test will cast doubt on a correctly estimated Sharpe ratio, but, again, can serve as a precautionary signal. However, we believe that both managers and investors tend to {\it underestimate} the length and depth of perfectly acceptable, ``normal'' drawdowns. Indeed, there is a 5\% chance for the drawdown of a ten-year process with $\textrm{SR}=0.5$ to last 7 years or more! Figures \ref{fig:conf} and \ref{fig:draw} are therefore helpful for managers and investors to draw a (fine) line between ``business as usual'' and possibly worrisome events.   

The proposed test allows one to update the Sharpe ratio estimate so that it becomes consistent with the drawdown observed. Let us assume, for example, that the manager determined that the drawdown depth $d_{\text{obs}}$ that she had observed is outside the 5\% interval, $d_{\text{obs}} > d_{5\%}$. She then can use Figure \ref{fig:draw} to update the Sharpe ratio to $\textrm{SR}_{\textrm{updated}}$ such that
\beq \label{d95}
\int_{d_{\text{obs}}}^{\infty} {\rm d}u\, \psi(u | {\textrm{SR}_{\textrm{updated}}}) = 0.05\,.
\eeq
In the same way, we may update the Sharpe ratio in order to render it compatible with the length of the observed drawdown. 

Finally, the knowledge of the joint distribution of $\ell$ and $d$ allows us to answer more complex questions. For example, having observed a drawdown of depth $d^*$, one can compute the conditional probability of the length $\ell$ of that drawdown. Let us define the boundaries of the 90\% probability region, $\ell^{\pm}(d^*)$, as
\beq
\mathbb{P}(\ell \geq \ell^{+} \,|\, d^*) = 0.05, \qquad \mathbb{P}(\ell \leq \ell^{-} \,|\, d^*) = 0.05.
\eeq
In other words, a typical drawdown of depth $d^*$ should last between $\ell^-$ and $\ell^+$ with 90 \% probability. We plot these boundary values for different values of Sharpe ratios in Figures \ref{fig:LD1}, \ref{fig:LD2}. In particular, we observe that a deep drawdown is unlikely to be very short. For example, for $\textrm{SR}=1.6$, the (unconditional) normalized 5 \% depth of a drawdown is $\approx 0.95$, but it is very unlikely that such a drawdown will end in less than $\ell^- \approx 2$ months.

\section{Acknowledgements}
We would like to thank Rapha\"{e}l Benichou, Emmanuel Buffet and Kay Wiese for discussions, and Paul Wilmott for suggesting to fully exploit the joint distribution of $\ell$ and $d$. 

\section{Appendix : The unconditional densities} \label{sec:app}
The maxima of stochastic processes have been extensively studied in mathematical literature. We shall not walk the reader through the history of this research, but rather immediately refer to \cite{mathpaper1, mathpaper2} and the references therein. These article prove an impressive result regarding the joint probability density of the value of the maximum (${B}^*_T$), time of the last visit time of the maximum ($\theta_T$) and the terminal value of the process ($B_T$) for various linear diffusion processes. The Brownian motion with drift is a linear diffusion process and the corresponding joint probability distribution may be found to be
\beqa \nn
F_{B_T, B_T^*, \theta_T}(a,b,s) &=& \frac{b(b-a)}{\pi (s(T-s))^{3/2}} \times \\ \nn
&&\textrm{exp} \left(\mu a -\tfrac{\mu^2 T}{2}-\tfrac{(b-a)^2}{2(T-s)} - \tfrac{b^2}{2s} \right)\,. \\
\eeqa
To adapt this function to the problem at hand, we perform the following change of variables
\beq
d := b - a, \qquad \ell:= T - s,
\eeq
which lends the depth of drawdown the status of an independent variable. The so-obtained $\widetilde F(d,b,\ell)$ is the unconditional joint probability density for the process to reach the maximum $b$ at time $T - \ell$ and subsequently suffer a drawdown of depth $d$.

Now, the quantities we used in the previous sections may be easily derived
\beqa
\rho(\ell) &=& \int^{\infty}_{-\infty} {\rm d}b \int^{\infty}_0 {\rm d}u \, \widetilde F(u,b,\ell)\,,\\
\psi(d) &=& \int^{\infty}_{-\infty}{\rm d}b \int^{T}_0 {\rm d}\ell \,  \widetilde F(d,b,\ell)\,.
\eeqa
The first density may be worked out analytically
\beqa \nn
\rho (\ell) &=& 2\, \left( \frac{1}{\sqrt{T - \ell}} \phi(\mu \sqrt{T - \ell}) + \mu \Phi(\mu \sqrt{T - \ell}) \right) \times \\ 
&&\left( \frac{1}{\sqrt{\ell}} \phi(\mu \sqrt{\ell}) - \mu \Phi(-\mu \sqrt{\ell}) \right) \,.
\eeqa
Here, $\phi(x)$ and $\Phi(x)$ are, respectively, the pdf and the cdf of the normal distribution. This result has been previously found in \cite{buffet} and \cite{jean-philippe}. 
The second density may be written as a single integral
\beqa \nn
\psi(d) &=& \frac{d}{\pi} \exp \left(-\mu\,d - \tfrac{\mu^2 T}{2} \right) \int^T_0 d\ell\, \frac{e^{ -\frac{d^2}{2\ell}}}{(\ell(T-\ell))^{3/2}} \times \\ \nn
&&\left(T - \ell + \sqrt{2\pi}\, \mu\, (T-\ell)^{\tfrac{3}{2}} \,e^{\tfrac{\mu^2 (T-\ell)}{2}} \,\Phi(\mu\sqrt{T - \ell})\right)\,.\\
\eeqa

\newpage


\begin{thebibliography}{99}
\bibitem{sigma} In the absence of the $\sigma=1$ normalization, the quantity entering the formulae of this paper would be the normalized depth of drawdown $d$ defined as the dollar amount of drawdown divided by the annualized dollar volatility.
\bibitem{fracBrown} On the last point, it would be very interesting to adapt the methods of Ref. \cite{Wiese} to compute the drawdown distribution of weakly correlated fractional Brownian motions.
\end{thebibliography}
\end{document}